\RequirePackage{lineno}
\documentclass[aps,prd,twocolumn,showpacs,superscriptaddress,groupedaddress,floatfix]{revtex4}  

\usepackage{graphicx}  
\usepackage{dcolumn}   

\usepackage{bm}        
\usepackage{amssymb}   
\usepackage{amsmath}
\usepackage{booktabs} 

\newcommand {\delm}{$\Delta m^{2}$}

\newcommand{\nus}{\nu_s}

\usepackage[normalem]{ulem}

\usepackage{color}
\usepackage{xspace, tabularx}
\usepackage{caption} 
\usepackage{threeparttable}

\newcommand{\beq}{\begin{equation}}
\newcommand{\eeq}{\end{equation}}
\newcommand{\beqa}{\begin{eqnarray}}
\newcommand{\eeqa}{\end{eqnarray}}
\newcommand{\n}{\noindent}
\newcommand{\nue}{\ensuremath{\nu_{e}}\xspace}
\newcommand{\numu}{\ensuremath{\nu_{\mu}}\xspace}

\begin{document}

\title{Search for active-sterile neutrino mixing using neutral-current interactions in NOvA
}
\hspace{5.2in} \mbox{FERMILAB-PUB-17-198-ND}

\newcommand{\ANL}{Argonne National Laboratory, Argonne, Illinois 60439, 
USA}
\newcommand{\IOP}{Institute of Physics, The Czech Academy of Sciences, 
182 21 Prague, Czech Republic}
\newcommand{\BHU}{Department of Physics, Institute of Science, Banaras 
Hindu University, Varanasi, 221 005, India}
\newcommand{\UCLA}{Physics and Astronomy Department, UCLA, Box 951547, Los 
Angeles, California 90095-1547, USA}
\newcommand{\Caltech}{California Institute of 
Technology, Pasadena, California 91125, USA}
\newcommand{\Cochin}{Department of Physics, Cochin University
of Science and Technology, Kochi 682 022, India}
\newcommand{\Charles}
{Charles University, Faculty of Mathematics and Physics,
 Institute of Particle and Nuclear Physics, Prague, Czech Republic}
\newcommand{\Cincinnati}{Department of Physics, University of Cincinnati, 
Cincinnati, Ohio 45221, USA}
\newcommand{\CSU}{Department of Physics, Colorado 
State University, Fort Collins, CO 80523-1875, USA}
\newcommand{\CTU}{Czech Technical University in Prague,
Brehova 7, 115 19 Prague 1, Czech Republic}
\newcommand{\Dallas}{Physics Department, University of Texas at Dallas,
800 W. Campbell Rd. Richardson, Texas 75083-0688, USA}
\newcommand{\Delhi}{Department of Physics and Astrophysics, University of 
Delhi, Delhi 110007, India}
\newcommand{\JINR}{Joint Institute for Nuclear Research,  
Dubna, Moscow region 141980, Russia}
\newcommand{\FNAL}{Fermi National Accelerator Laboratory, Batavia, 
Illinois 60510, USA}
\newcommand{\UFG}{Instituto de F\'{i}sica, Universidade Federal de 
Goi\'{a}s, Goi\^{a}nia, Goi\'{a}s, 74690-900, Brazil}
\newcommand{\Guwahati}{Department of Physics, IIT Guwahati, Guwahati, 781 
039, India}
\newcommand{\Harvard}{Department of Physics, Harvard University, 
Cambridge, Massachusetts 02138, USA}
\newcommand{\IHyderabad}{Department of Physics, IIT Hyderabad, Hyderabad, 
502 205, India}
\newcommand{\Hyderabad}{School of Physics, University of Hyderabad, 
Hyderabad, 500 046, India}
\newcommand{\Indiana}{Indiana University, Bloomington, Indiana 47405, 
USA}
\newcommand{\INR}{Inst. for Nuclear Research of Russia, Academy of 
Sciences 7a, 60th October Anniversary prospect, Moscow 117312, Russia}
\newcommand{\Iowa}{Department of Physics and Astronomy, Iowa State 
University, Ames, Iowa 50011, USA}
\newcommand{\Irvine}{Department of Physics and Astronomy, 
University of California at Irvine, Irvine, California 92697, USA}
\newcommand{\Jammu}{Department of Physics and Electronics, University of 
Jammu, Jammu Tawi, 180 006, Jammu and Kashmir, India}
\newcommand{\Lebedev}{Nuclear Physics Department, Lebedev Physical 
Institute, Leninsky Prospect 53, 119991 Moscow, Russia}
\newcommand{\MSU}{Department of Physics and Astronomy, Michigan State 
University, East Lansing, Michigan 48824, USA}
\newcommand{\Duluth}{Department of Physics and Astronomy, 
University of Minnesota Duluth, Duluth, Minnesota 55812, USA}
\newcommand{\Minnesota}{School of Physics and Astronomy, University of 
Minnesota Twin Cities, Minneapolis, Minnesota 55455, USA}
\newcommand{\Oxford}{Subdepartment of Particle Physics, 
University of Oxford, Oxford OX1 3RH, United Kingdom}
\newcommand{\Panjab}{Department of Physics, Panjab University, 
Chandigarh, 106 014, India}
\newcommand{\RAL}{Rutherford Appleton Laboratory, Science and 
Technology Facilities Council, Didcot, OX11 0QX, United Kingdom}
\newcommand{\SAlabama}{Department of Physics, University of 
South Alabama, Mobile, Alabama 36688, USA} 
\newcommand{\Carolina}{Department of Physics and Astronomy, University of 
South Carolina, Columbia, South Carolina 29208, USA}
\newcommand{\SDakota}{South Dakota School of Mines and Technology, Rapid 
City, South Dakota 57701, USA}
\newcommand{\SMU}{Department of Physics, Southern Methodist University, 
Dallas, Texas 75275, USA}
\newcommand{\Stanford}{Department of Physics, Stanford University, 
Stanford, California 94305, USA}
\newcommand{\Sussex}{Department of Physics and Astronomy, University of 
Sussex, Falmer, Brighton BN1 9QH, United Kingdom}
\newcommand{\Tennessee}{Department of Physics and Astronomy, 
University of Tennessee, Knoxville, Tennessee 37996, USA}
\newcommand{\Texas}{Department of Physics, University of Texas at Austin, 
Austin, Texas 78712, USA}
\newcommand{\Tufts}{Department of Physics and Astronomy, Tufts University, Medford, 
Massachusetts 02155, USA}
\newcommand{\UCL}{Physics and Astronomy Dept., University College London, 
Gower Street, London WC1E 6BT, United Kingdom}
\newcommand{\Virginia}{Department of Physics, University of Virginia, 
Charlottesville, Virginia 22904, USA}
\newcommand{\WSU}{Department of Mathematics, Statistics, and Physics,
 Wichita State University, 
Wichita, Kansas 67206, USA}
\newcommand{\WandM}{Department of Physics, College of William \& Mary, 
Williamsburg, Virginia 23187, USA}
\newcommand{\Winona}{Department of Physics, Winona State University, P.O. 
Box 5838, Winona, Minnesota 55987, USA}
\newcommand{\Crookston}{Math, Science and Technology Department, University 
of Minnesota -- Crookston, Crookston, Minnesota 56716, USA}
\newcommand{\deceased}{Deceased.}

\affiliation{\ANL}
\affiliation{\IOP}
\affiliation{\BHU}
\affiliation{\Caltech}
\affiliation{\Charles}
\affiliation{\Cincinnati}
\affiliation{\Cochin}
\affiliation{\CSU}
\affiliation{\CTU}
\affiliation{\Delhi}
\affiliation{\FNAL}
\affiliation{\UFG}
\affiliation{\Guwahati}
\affiliation{\Harvard}
\affiliation{\Hyderabad}
\affiliation{\IHyderabad}
\affiliation{\Indiana}
\affiliation{\INR}
\affiliation{\Iowa}
\affiliation{\Irvine}
\affiliation{\Jammu}
\affiliation{\JINR}
\affiliation{\Lebedev}
\affiliation{\MSU}
\affiliation{\Duluth}
\affiliation{\Minnesota}
\affiliation{\Panjab}
\affiliation{\SAlabama}
\affiliation{\Carolina}
\affiliation{\SDakota}
\affiliation{\SMU}
\affiliation{\Stanford}
\affiliation{\Sussex}
\affiliation{\Tennessee}
\affiliation{\Texas}
\affiliation{\Tufts}
\affiliation{\UCL}
\affiliation{\Virginia}
\affiliation{\WSU}
\affiliation{\WandM}
\affiliation{\Winona}

\author{P.~Adamson}
\affiliation{\FNAL}


\author{L.~Aliaga}
\affiliation{\FNAL}

\author{D.~Ambrose}
\affiliation{\Minnesota}


\author{N.~Anfimov}
\affiliation{\JINR}


\author{A.~Antoshkin}
\affiliation{\JINR}

\affiliation{\Minnesota}

\author{E.~Arrieta-Diaz}
\affiliation{\SMU}

\author{K.~Augsten}
\affiliation{\CTU}

\author{A.~Aurisano}
\affiliation{\Cincinnati}


\author{C.~Backhouse}
\affiliation{\Caltech}

\author{M.~Baird}
\affiliation{\Sussex}
\affiliation{\Indiana}

% ...

\collaboration{The NOvA Collaboration}
\noaffiliation

\date{\today}

\begin{abstract}
We report results from the first search for sterile neutrinos mixing with active neutrinos through a reduction in the rate of neutral-current interactions over a baseline of 810~km between the NOvA detectors.
Analyzing a 14-kton detector equivalent exposure of 6.05$\times$10$^{20}$ protons-on-target in the NuMI beam at Fermilab, we observe 95 neutral-current candidates at the Far Detector compared with $83.5 \pm 9.7 \mbox{(stat.)} \pm 9.4 \mbox{(syst.)}$ events predicted assuming mixing only occurs between active neutrino species. No evidence for $\nu_{\mu} \rightarrow \nu_{s}$ transitions is found. Interpreting these results within a 3+1 model, we place constraints on the mixing angles $\theta_{24}$\,$<$\,$20.8^{\circ}$ and $\theta_{34}$\,$<$\,$31.2^{\circ}$ at the 90\% C.L.  for $0.05$~eV$^2\leq \Delta m^2_{41}\leq 0.5$~eV$^2$, the range of mass splittings that produce no significant oscillations over the Near Detector baseline.

\end{abstract}

\maketitle

Mixing between the three known active neutrinos \numu, \nue, and $\nu_\tau$ has been well established by measurements of neutrinos produced in a variety of sources, including neutrinos created in the Earth's atmosphere, in the Sun, in accelerators, and in terrestrial reactors~\cite{ref:atmos2_1,ref:sno_2,ref:kamland_2, ref:K2K, ref:minosdisappear_4,ref:T2K, ref:dayabay, ref:reno,ref:doublechooz,ref:opera, ref:novaprls}. 
However, additional neutrino flavors that mix with the active flavors may exist. If indeed there is a fourth neutrino mass eigenstate in addition to the states $\nu_1$, $\nu_2$, and $\nu_3$, a new linearly independent state can be formed
\begin{equation}\label{eq:super}
|\nu_{s}\rangle = \sum^{4}_{i=1}U^*_{s i} |\nu_{i}\rangle\,,
\end{equation}
where $U$ represents a unitary $4\times4$ extended Pontecorvo-Maki-Nakagawa-Sakata matrix~\cite{ref:PMNS1, ref:PMNS2}, and the $\nu_i$ denote the mass eigenstates. This $\nu_s$ neutrino would not have a standard model charged lepton partner, so it could not couple to the $W$ boson. Further, LEP measurements of the invisible decay of the $Z^0$ boson~\cite{ref:pdg} are consistent with three neutrino flavors implying that any additional neutrino state $\nu_s$ is either very massive or it is {\it sterile} and does not participate in the weak interaction~\cite{ref:sterile}. 
Identical arguments can be applied to scenarios with two or more $\nu_s$ states. The discovery of a new sterile neutrino state with a mass below half the $Z^0$ boson mass could help explain the smallness of neutrino masses~\cite{ref:zhang_seesaw}. In addition, $\nu_s$'s are also dark matter candidates, as they may have a wide range of masses and have no mechanism to directly decay into lighter particles over time scales comparable to the age of the Universe due to their absence of nongravitational interactions with matter~\cite{ref:sterile}. 
Furthermore, $\nu_s$'s may explain puzzling questions related to the fusion reaction rate during core-collapse supernovae~\cite{ref:sterile}.
Data from the short-baseline experiments LSND and MiniBooNE~\cite{ref:lsnd,ref:miniboone} are compatible with active-sterile neutrino oscillations driven by a new \delm\,of the order of $1$~eV$^2$, but this evidence is inconclusive~\cite{ref:karmen,ref:mbunumu}. 
A deficit of \nue~consistent with the same \delm~range has been observed in measurements with calibration sources used by the SAGE and GALLEX gallium experiments~\cite{ref:Acero-Ortega, ref:Giunti}. Several other short-baseline and long-baseline searches have found no evidence for these light $\nu_s$ states and place strong constraints on their existence~\cite{ref:minosdayabay, ref:minos, ref:dayabaysterile, ref:superk,ref:IceCube}.
Meanwhile, calculations of reactor $\bar{\nu}_e$ fluxes \cite{ref:Huber, ref:Mueller} predict a value 3\% larger on average than experiments have observed, which has been interpreted as $\bar{\nu}_e$ disappearance; but recent measurements of the reactor core fuel evolution \cite{PhysRevLett.118.251801}, and observation of spectral distortions independent of distance to reactor cores \cite{PhysRevLett.118.042502}, disfavor this interpretation.\\
The NOvA experiment can search for oscillations into $\nu_s$'s by looking for disappearance of the active neutrino flux between the Near Detector (ND) and Far Detector (FD). In the analysis presented here, we focus on the neutral-current (NC) channel. Oscillations into a fourth light $\nu_s$ state would result in an energy-dependent suppression of the NC event rate, as the $\nu_s$ would not interact in the detector. This suppression contrasts with the effects of standard oscillations among the three active neutrinos, which leave the NC rate and spectrum unchanged. 
This paper presents the first NOvA results from a search for light $\nu_s$ mixing by looking for a depletion of the NC event rate at the FD with respect to the prediction derived from ND observations.\\
The NOvA experiment consists of the Far and Near Detectors,  placed  810~km and 1~km from Fermilab's NuMI beam source~\cite{ref:NUMI}, respectively. The FD is located on the surface in northern Minnesota, 14.6~mrad off the beam axis, and the ND is located at Fermilab 100~m underground and samples the same off-axis angle as the FD, ensuring similarity in the energy spectra observed at the two detectors. 
The NuMI neutrino beam is produced using 120\,GeV protons incident on a 1.2~m-long graphite target. The kaons and pions emerging from the target are focused by two magnetic horns and either decay in flight into neutrinos over a distance of 705~m, including a 675~m decay pipe, or are absorbed.
The resulting neutrino beam has a narrow energy spectrum, with a full width half maximum of approximately 1~GeV peaked at 2~GeV. The ND sees a larger solid angle as it is closer to the beam source, and hence a wider energy distribution.  The beam is extracted for $10~\mu$s every 1.33~s and is composed primarily of $\nu_\mu$. Simulation predicts small contaminations of 1.8\% $\bar{\nu}_\mu$ and 0.7\% $\nu_e$\,+\,$\bar{\nu}_e$ in the 1$-$3~GeV energy range. 

NOvA's design provides several distinct advantages over other long-baseline neutrino experiments for probing sterile neutrinos through the NC channel. The fully active detector technology offers superior reconstruction, identification and energy determination of NC events. In addition, the  narrow-band beam centered at the three-flavor oscillation maximum results in a large expected NC signal with significantly reduced backgrounds providing excellent sensitivity to the $\theta_{34}$ mixing angle, as described in detail below.

The two detectors are functionally identical tracking calorimeters, composed of cells filled with a mineral oil-based liquid scintillator doped with 5\% pseudocumene~\cite{OIL}. The cells are 3.9 by 6.6~cm constructed from reflective PVC~\cite{PVC}. 
The scintillator accounts for 62\% of the detector mass. The FD (ND) cells are 15.5 (3.9)~m long and contain a loop of wavelength-shifting fiber with both ends read out by one pixel of a 32-pixel Hamamatsu avalanche photodiode. A total of 344,064 (18,432) cells are organized into 896 (192) planes arranged so that the cells alternate between horizontal and vertical orientations, relative to the beam axis, to enable three-dimensional reconstruction. The FD and ND have masses of 14~kt and 193~t, respectively.
The FD is covered by a  3~m overburden of concrete and barite which blocks most of the electromagnetic and hadronic components of cosmic ray secondaries. 
Pulse height and timing for all energy deposits above a preset threshold are read out in a 550~$\mu$s window centered around the 10~$\mu$s beam spill. In addition, there is a 550~$\mu$s minimum-bias trigger run at 10~Hz to provide a high-statistics cosmogenic background sample.

This analysis uses data collected from February 2014 to May 2016, corresponding to beam powers ranging between 250 and 560~kW, and including periods of partial-detector operation. During this time, the experiment collected 6.68\,$\times$\,$10^{20}$\,protons-on-target (POT), 
equivalent to a full-detector exposure of 6.05\,$\times$\,$10^{20}$\,POT. 

We simulate neutrinos resulting from decays of mesons produced by proton interactions in the NuMI beam target using the {\sc FLUKA}~\cite{ref:fluka1,ref:fluka2} simulation package and the {\sc FLUGG}~\cite{ref:flugg} {\sc Geant4} geometry interface. Neutrino interactions in the detector and the  surrounding material are modeled by passing the simulated flux to {\sc GENIE}~\cite{ref:GENIE}. {\sc Geant4}~\cite{ref:geant1, ref:geant2} propagates the resulting particles through the detector to determine the energy deposited in the active material. A custom simulation models the propagation of photons in the detector cells, the light attenuation in the fibers, and  the response of the APDs and the front-end electronics~\cite{ref:simChain}.

The first step in the reconstruction of neutrino interactions is the clustering of energy deposits close together in space and time, as they are likely to be associated with a single interaction~\cite{ref:eventReco}. These clusters form the event to be reconstructed.
The energy response of the detector is calibrated using cosmic ray muons, which are used to set the absolute energy scale, as well as to determine a correction for attenuation along the wavelength-shifting fibers. 
We define the calorimetric energy of an event as the sum of calibrated energy deposits of the cluster.
To reconstruct individual particles within an event, a Hough transform~\cite{ref:Hough} is applied to the cluster and a three-dimensional vertex is determined from a fit to the resulting lines' most likely common origin.  The spatial locations of energy deposits are clustered around the vertex into prongs (clusters with defined starting point and direction), each containing deposits attributed to a final-state particle.

In NC neutrino interactions in the NOvA detectors, where a $Z^{0}$ boson is exchanged primarily with a carbon nucleus, the neutrino leaves the detector with reduced energy and products of nuclear fragmentation remain behind. This hadronic recoil appears in the detector as an isolated cluster of energy deposits, distinguishable from the charged-current (CC) interactions by the lack of a charged track, or compact energy deposit, associated with the lepton.
Backgrounds arise from both misidentified CC neutrino interactions and from external sources.
NuMI beam \numu\,CC and \nue\,CC events, typically with high momentum transfer to the hadronic system, can be produced where the lepton may be misidentified or not reconstructed, thus mimicking a NC neutrino interaction. Backgrounds due to $\nu_{\tau}$\,CC events are found to be negligible.
External events are primarily cosmogenic neutrons produced in the FD overburden, and NuMI beam events interacting in the periphery of the ND and in the surrounding cavern. 
The predicted proportions of different event types differ substantially between the two detectors: $\nu_{\mu}$~CC ($\nu_{e}$~CC) interactions at the FD are suppressed (enhanced) by oscillations as compared to the ND. 
On average, before applying additional selections, we reconstruct 74,000 cosmogenic events for each reconstructed neutrino event in the 10~$\mu$s beam spill window at the FD. As the ND is located underground, cosmogenic backgrounds are negligible at the ND.

All events are required to have a reconstructed vertex and at least one reconstructed prong that spans a minimum of two detector planes. The entirety of the prong is required to be at least 10~cm (25~cm) away from the FD (ND) walls.
The events which pass these selections are additionally required to have a calorimetric energy between 0.5 and 4~GeV. This criterion rejects low-energy events, where combined uncertainties in energy resolution and threshold are substantial, and avoids higher-energy regions where the ND and FD selection efficiencies diverge due to the smaller size of the ND. 

\begin{figure}
	\includegraphics[scale=0.45]{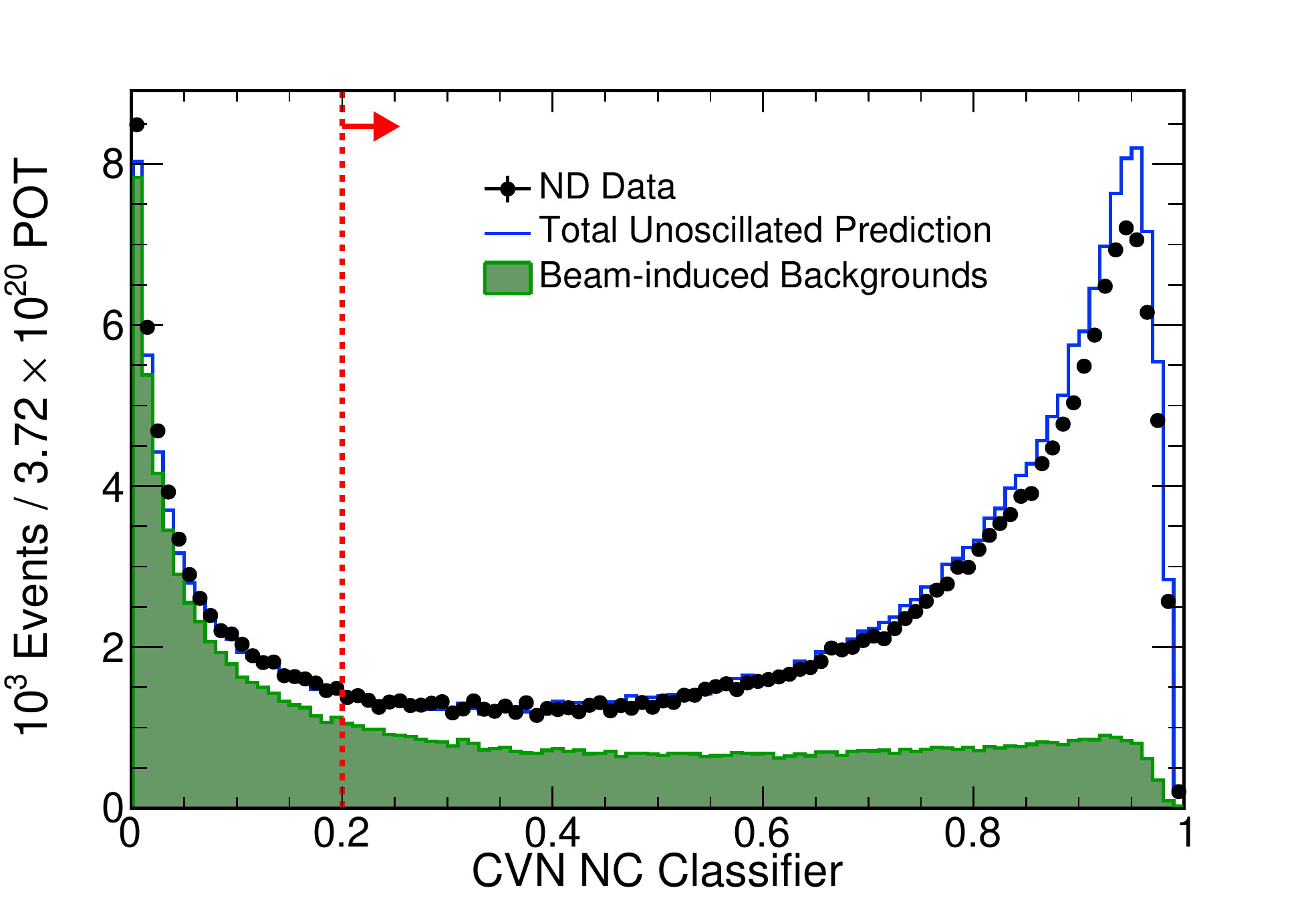}
    \captionsetup{justification=RaggedRight}
	\caption{The CVN NC classifier for ND data and simulation. The beam-induced backgrounds are \numu CC and \nue CC events originating both internally and externally to the detector. }
	\label{NDCVN}
\end{figure}

To separate beam NC neutrino interactions from beam CC neutrino and cosmogenic interactions, we use a convolutional neural network algorithm, based on a modified GoogLeNet architecture~\cite{ref:googlenet}. This algorithm, the Convolutional Visual Network (CVN)~\cite{ref:CVN}, extracts classification features using a series of transformations to the pattern of energy deposits within the detector, and then uses these features to determine the likelihood that a particle interaction is of a particular type. The CVN algorithm simultaneously provides classifiers for multiple particle types, giving it general applicability within NOvA. For example, the CVN $\nu_e$ CC classifier has been used as the primary selector in the most recent NOvA $\nu_e$ appearance analysis~\cite{ref:novanueSA}. The CVN NC classifier is used in this analysis to separate the NC signal from backgrounds, and the distribution of likelihoods resulting from its application to ND data and simulation is shown in Fig.~\ref{NDCVN}.

FD cosmogenic background rejection is optimized using a high-statistics minimum-bias cosmic data sample. In addition to the CVN selection, we apply the following criteria: 
to remove cosmogenic neutron backgrounds in the FD, the reconstructed start and end position of prongs must be a minimum distance of 5~m away from the top of the detector; to remove downward-going cosmogenic activity, the fractional transverse momentum, with respect to the beam direction, of the highest energy prong is required to be less than 0.8; and, finally, to remove the remaining contained cosmogenic backgrounds, a boosted decision tree is employed~\cite{ref:novanumuFA}. After all selections, the effective fiducial masses of the FD and ND are 8.83~kt and 34~t, respectively.
The cosmogenic background rate is estimated from NuMI-triggered data, excluding a 30~$\mu$s window centered on the beam spill. This sample reproduces the detector configuration and quality conditions of the data within the beam spill. A rejection level where only 1 in every 1.7 million cosmogenic events is misidentified as a NC signal event is obtained, equivalent to 1 cosmogenic event every 60,600 spills. 

At the FD (ND), we achieve a 50\% (62\%) NC signal efficiency and  72\% (70\%) NC signal purity for contained events within the fiducial volume. This selection results in $173,000$ selected ND data events, with a predicted background of 53,700 \numu CC and 1,700 \nue CC events.

\begin{figure}
    \includegraphics[scale=0.45]{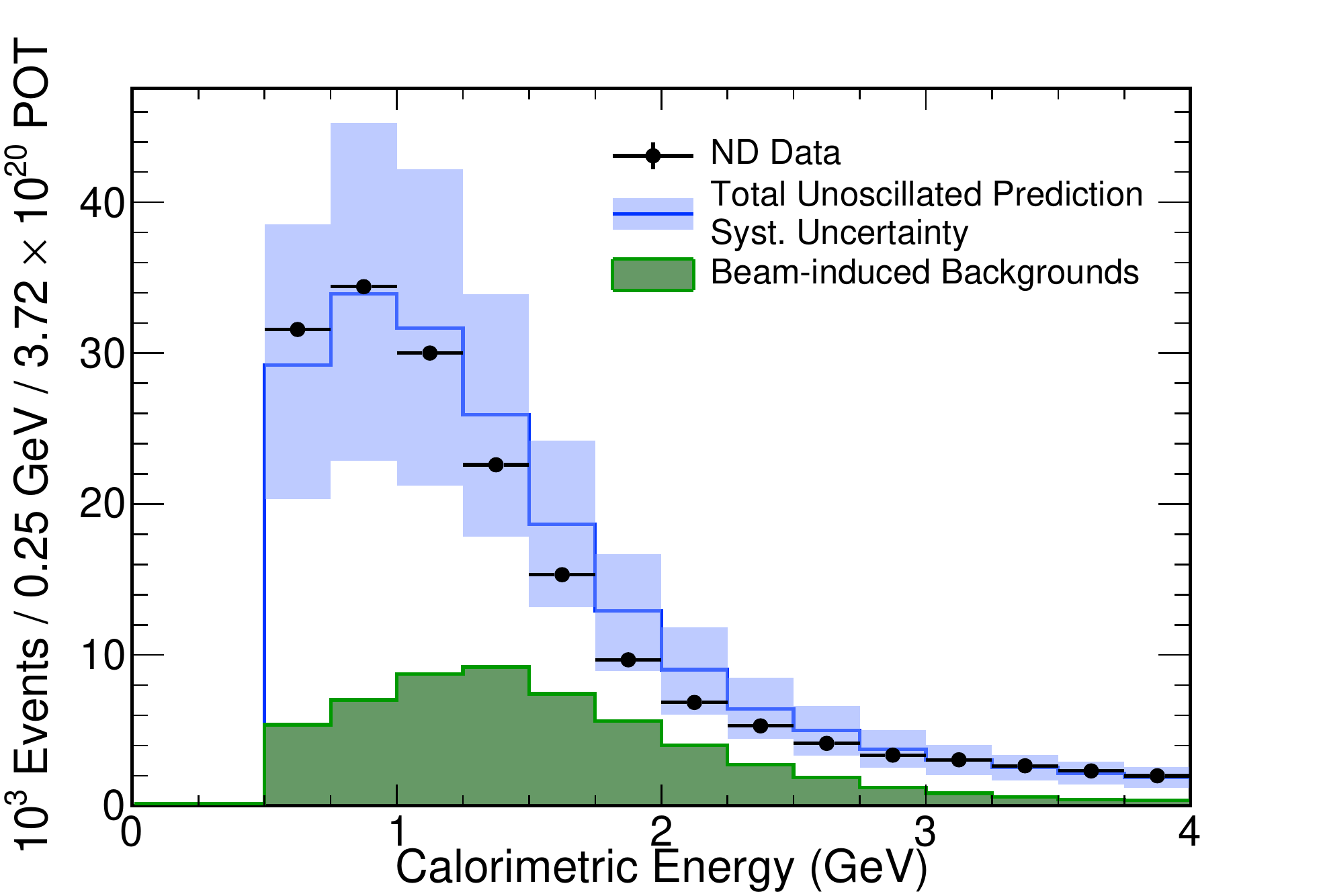}
    \captionsetup{justification=RaggedRight}
	\caption{The unoscillated calorimetric energy spectrum for NC selected data and simulated events at the ND. The beam-induced backgrounds are \numu CC and \nue CC events originating both internally and externally to the detector.}
	\label{NDecal}
\end{figure}
Our search for active-sterile neutrino oscillations proceeds by comparing the predicted rate in the FD with the observed NC events in the selected calorimetric energy range.
Though no spectral shape information is directly used for this comparison at the FD, the FD rate prediction does have a dependence on the ND calorimetric energy shape through our extrapolation procedures, as discussed below.
The FD rate is predicted from the calorimetric energy spectrum for NC-selected events in the ND. The comparison of the ND spectra in data and simulation reveals discrepancies attributable to limitations in the simulation and detector response modeling. 
Results from $\nu_{\mu}$ CC measurements in NOvA~\cite{ref:numuSA} and MINERvA~\cite{ref:minvera2016MEC} indicate that there are unmodeled nuclear effects in {\sc GENIE} (2.10.2) at low hadronic
recoil energy, caused by scattering of neutrinos from correlated nucleon pairs within the nucleus \cite{ref:theory1, ref:theory2,ref:bubblechamber1, ref:theory4}.  A parallel process is expected to result in similar NC interactions, which would also be unmodeled in the simulation.
The energy threshold required ensures these have a minimal effect on this analysis. 
An excess in the simulation rate is seen at higher hadronic recoils, consistent with measurements of $\nu_\mu$~CC($\pi^+$) from the MINERvA experiment~\cite{ref:minerva2016numupi}, which observed a data rate 1$-$2~$\sigma$ below simulation. Improved agreement with the ND data was achieved by applying a 35\% reduction in CC and NC deep inelastic scattering events with final-state invariant mass, $W$, less than 1.7~GeV. This reduction models the nonresonant single pion overproduction in {\sc GENIE} suggested by a recent reanalysis of $\nu_\mu$-deuterium pion production data \cite{ref:bubblechamber1, ref:bubblechamber2}. The calorimetric energy spectra obtained from data and simulation after this correction are displayed in Fig.~\ref{NDecal}.

The differences observed between the ND data and simulation are mainly accounted for by our FD prediction technique, which extrapolates the observed ND spectra to the FD while accounting for flux and acceptance differences as calculated from the simulation.
Any remaining data-simulation differences are absorbed within systematic uncertainties. Furthermore, we perform a rate-only measurement to ensure the analysis is negligibly affected by the potentially absent components of the simulation modeling described above. 
This analysis restricts itself to a $\nu_{s}$ mass range that does not induce oscillations within the ND baseline.

Since the NC signal, and the \numu~CC and \nue~CC backgrounds, are subject to distinct oscillation probabilities, they are extrapolated separately to the FD. The observed ND spectrum is decomposed into NC, \numu~CC, and \nue~CC components based on the proportion of each component predicted in the simulation per 0.25~GeV calorimetric energy bin. This decomposition distributes the observed ND discrepancies between the data and simulation among all interaction modes based on their simulated proportional contribution per bin. These ND components are then converted to true neutrino energy bins using simulated migration matrices.

To obtain the predicted NC-selected FD spectrum, $F^{\rm{pred}}$, we apply a far/near ratio extrapolation procedure. As described by Eq.~(\ref{eq:FoverN}), for each true interaction type $k$ $\in$ \{NC, CC\} and neutrino flavor $\nu_{\alpha}$, the ratio of ND NC-selected data and simulation, $N_{jk\alpha}^{\rm{data}}/N_{jk\alpha}^{\rm{sim}}$, is used to correct the FD NC-selected simulated true energy spectrum $F_{jk\beta}^{\rm{sim}}$ in true energy bins $j$. These FD spectrum bins are multiplied by the relevant oscillation probabilities $P(\nu_{\alpha},\nu_{\beta})$ computed in true energy, to obtain

\begin{equation}   \displaystyle  
 F_{jk\beta}^{\rm{pred}} = \sum_{\alpha}\frac{N_{jk\alpha}^{\rm{data}}}{N_{jk\alpha}^{\rm{sim}}} 
\displaystyle
~F_{jk\beta}^{\rm{sim}} ~P(\nu_{\alpha},\nu_{\beta}).
  \label{eq:FoverN}
\end{equation} 
The $F_{jk\beta}^{\rm{pred}}$ are then translated from true energy bins into bins of calorimetric energy, using simulated migration matrices for each interaction type, $k$, and flavor after oscillation, $\beta$. The predictions for each component are summed together and integrated over bins of calorimetric energy. Finally, the result is summed with the cosmogenic background, and the negligible $\nu_\tau$ CC background, estimated from simulation, to provide the predicted FD event rate $F^{\rm{pred}}$. 

Systematic uncertainties on the rate of NC events in the FD are evaluated, one parameter at a time, by generating sets of modified simulated events that are propagated through the full extrapolation and analysis chain to produce shifted FD predictions. Any difference in the prediction from nominal is taken as the systematic uncertainty. Many sources of systematic uncertainty are highly correlated between the two functionally identical detectors. Absolute uncertainties, defined as uncertainties that affect both detectors in the same way, largely cancel in this analysis. However, we also take into account relative uncertainties, specific to either one of the detectors, that do not cancel, resulting in the largest contributions to the overall systematic error. The systematic uncertainties are summarized in Table~\ref{tab:SystSummary}.
 
 \begin{table*} [!ht]
  \begin{centering}
  \begin{threeparttable} 
  \captionsetup{justification=RaggedRight}
  \caption[Systematic Uncertainty Summary]{The effect of the systematic uncertainties on the NC and CC expected event rates, and on the sensitivity to $\theta_{24}$ and $\theta_{34}$. For the systematic uncertainties on the rates, the total is the sum of the absolute individual uncertainties added in quadrature, whereas the total systematic effect on the mixing angles is calculated with all sources of uncertainty applied simultaneously. In all cases, the illustrative effects shown for each individual absolute uncertainties are calculated independently.}
    \begin{tabular}{lc c c c c c lc}
      \hline\hline
      & NC signal & CC background & Effect on & Effect on \\
      Source of uncertainty & difference (\%) & difference (\%) & $\theta_{24}$ limit ($\%$) & $\theta_{34}$ limit ($\%$) \\
      \hline
      ND composition & 7.0 & 10.4 & 7.5 & 7.4 \\
      Calibration & 5.8 & 6.0 & 6.4 & 7.3 \\
      Normalization & 4.9 & 4.9 & 4.6 & 4.6 \\
      ND external activity & 4.1 & 1.7 & 2.9 & 2.3 \\
      Beam flux & 3.4 & 3.6 & 0.6 & 0.8 \\
      Scintillation model & 2.4 & 1.8 & $<$0.1 & $<$0.1 \\
      Simulation statistics & 2.0 & 4.8 & 1.2 & 1.2 \\
      Neutrino interaction & 1.6 & 4.8 & $<$0.1 & $<$0.1 \\
      Acceptance & 1.0 & 0.6 & $<$0.1 & $<$0.1 \\
      Three-flavor oscillation parameters & 0.7 & 10.7 & $<$0.1 & $<$0.1 \\\\
      Total  & 12.2 & 15.3 & 22.0 & 21.7 \\
      \hline\hline
    \end{tabular}
    \label{tab:SystSummary}
    \end{threeparttable}
  \end{centering}
\end{table*}

The dominant source of systematic uncertainty is attributed to a mismodeling of either the NC signal or the CC background rates observed in the ND.
To assess the size of this uncertainty, the extrapolation procedure is carried out with the entirety of the 
observed ND data-simulation difference attributed either to the NC signal or the $\nu_\mu$ CC background while simultaneously assuming a 100\% scale uncertainty on the small intrinsic beam $\nu_e$ CC component. 
The former results in a reduction of the predicted NC-signal sample at the FD when compared to the nominal FD prediction  (NC-signal and CC-background are both allowed to vary).  
The latter results in an increase of the number of predicted NC events at the FD compared to the nominal prediction. This change from nominal is larger than when assigning the excess exclusively to $\nu_\mu$ CC events, as these are suppressed at the FD by three-flavor oscillations.
We assign a 7.0\% uncertainty on the NC signal and a 10.4\% uncertainty on the CC backgrounds to account for this difference.

A 5\% uncertainty on the absolute and relative calibrations between the detectors is determined through the observed data-simulation differences in several probes including Michel electrons and the measured $\pi^0$ mass peak. As these probes are only studied in the ND, this uncertainty is conservatively applied as both an absolute and relative uncertainty. 
This leads to a 5.8\% uncertainty on the NC signal and a 6.0\% uncertainty on the CC backgrounds in the FD, arising from threshold selection effects and changes in the selection efficiency with energy. 

A normalization systematic of 4.9\% is estimated for both the NC signal and CC backgrounds. The dominant contributions arise from a 3.7\% difference between simulated FD neutrino interactions with and without overlaid minimum-bias cosmogenic data and a 2.9\% uncertainty from the ND data-simulation differences in prong reconstruction.These effects are both due to reconstruction inefficiencies due to multiple interactions in the detector per beam pulse.  Other subpercent contributions include the uncertainties on the detector noise model, the mass of the detector, the POT counting, and the variation of the beam intensity. 

Uncertainties on the cross section and hadronization models used for the predictions are calculated using the {\sc GENIE} event reweighting framework~\cite{ref:GENIE_MANUAL}.
In addition, a 50\% uncertainty on the normalization of the {\sc GENIE} component modeling of CC scattering from correlated nucleons is included, motivated by the data/simulation discrepancies seen in the $\nu_\mu$-CC channel~\cite{ref:numuSA}. Further, the full size of the 35\% scaling applied to deep inelastic scattering events with $W$\,$<$\,1.7~GeV is included as an uncertainty. This leads to a 1.6\% uncertainty on the NC signal and a 4.8\% uncertainty on the CC backgrounds in the FD.

\begin{figure}\includegraphics[scale=0.45]{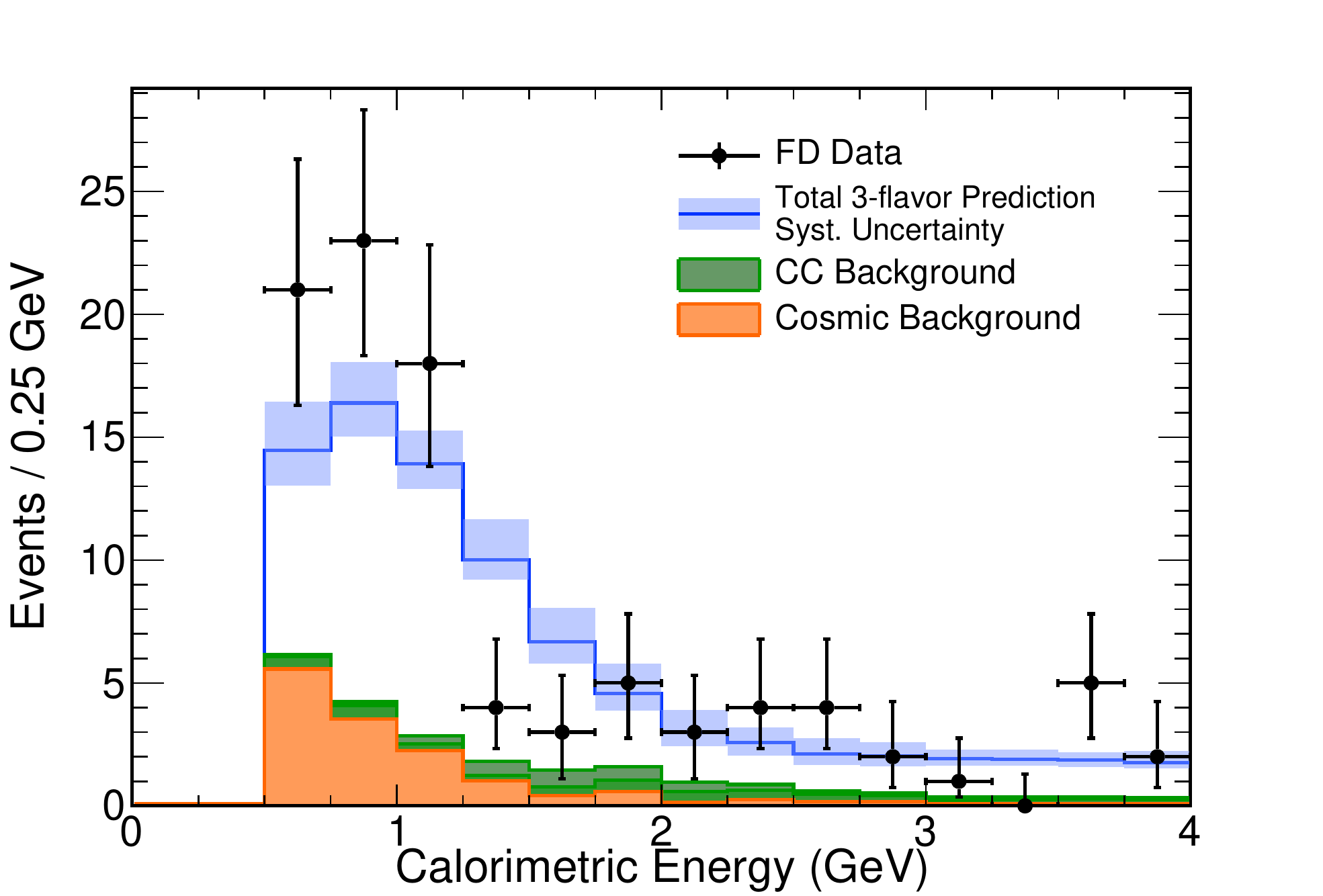}
\captionsetup{justification=RaggedRight}
\caption{The three-flavor FD calorimetric energy spectrum for NC selected data and predicted events for the 6.05$\times10^{20}$\,POT-equivalent.}
\label{FD_cale}
\end{figure}

Other less significant sources of systematic uncertainties include the beam flux model, the modeling of scintillator response, the effect of using limited statistics for the simulation, the possible contamination of the ND spectrum by events originating in materials outside of the detector, and potential mismodeling of acceptance differences between the ND and FD due to their differing sizes. A shift of the three-flavor oscillation parameters by the 1~$\sigma$ deviations from their nominal values~\cite{ref:pdg} changes the FD prediction by no more than a single event. This effect is also included as a systematic uncertainty.
The sum in quadrature of all effects results in a 12.2\% uncertainty on the NC signal and a 15.3\% uncertainty on the CC backgrounds.\\
Upon examining the FD data, 95~NC~event candidates are observed, with ${83.5\,\pm\,9.7 \mbox{(stat.)} \pm 9.4 \mbox{(syst.)}
}$ events predicted under the three-flavor oscillation assumption. Values for $\theta_{12}$, $\theta_{13}$, $\theta_{23}$, $\Delta m^{2}_{21}$, and $\Delta m^{2}_{32}$ are taken from~\cite{ref:pdg}, with normal hierarchy and maximal mixing assumed.  Matter effects are included in the oscillation probability calculations, with the Earth's crust density assumed to be uniformly 2.84\,g/cm$^3$~\cite{ref:density}. The value of $\delta_{\rm CP}$ is set to 0, as its effect is negligible. Table \ref{tab:extrap} shows the breakdown of the predicted events in the FD and Fig.~\ref{FD_cale} shows the calorimetric energy distribution of the selected data events in the FD under the three-flavor model assumption.

\begin{table*}[!ht]
  \begin{centering}
  \begin{threeparttable}
  \captionsetup{justification=RaggedRight}
  \caption{Extrapolated prediction of FD event counts normalized to $6.05\times 10^{20}$ POT. The systematic (statistical) uncertainty is shown for signal and background (cosmogenic events).}
  \label{tab:extrap}
  \begin{tabularx}{.8\textwidth}{X >{\centering\arraybackslash}X >{\centering\arraybackslash}X >{\centering\arraybackslash}X >{\centering\arraybackslash}X >{\centering\arraybackslash}X}
    \hline\hline
     &  & \multicolumn{3}{c}{CC background} & \\\cline{3-5}
     Total & NC signal &  $\nu_{\mu}$ & $\nu_{e}$  & $\nu_{\tau}$ & Cosmics\\
    \hline \\
      83.5$\pm$9.4 & 60.6$\pm$7.4 &  4.6$\pm$0.7 & 3.6$\pm$0.6 & 0.4$\pm$0.1 & 14.3$\pm$0.7\\
    \hline\hline
  \end{tabularx}
   \end{threeparttable}
 \end{centering}
\end{table*}
The statistic $R_{\rm NC}$~\cite{ref:minosNCPRL} is computed as a model independent test for active to sterile mixing,
\beq R_{\rm NC} \equiv \frac{
F^{\rm{data}} - \sum F^{\rm{pred}} ({\rm{bkg})}}{
F^{\rm{pred}}({\rm NC})}, \label{eq:R} \eeq
\n where the predicted quantities are calculated assuming three-flavor oscillations. 

Active to sterile mixing would reduce $F^{\rm{data}}$ relative to the three-flavor signal component $F^{\rm{pred}}$ (NC) and the sum of the multiple background components $\sum{F^{\rm{pred}}\rm{(bkg)}}$, both derived from the total FD prediction $F^{\rm{pred}}$ described in Eq.~(\ref{eq:FoverN}), resulting in $R_{\rm NC} < 1$.
We measure $R_{\rm NC} = 1.19 \pm 0.16 \mbox{(stat.)} {+0.10} \mbox{(syst.)}$, corresponding to a 1.03\,$\sigma$ excess over the three-flavor prediction of $R_{\rm NC}$\,=\,1, and consistent with three-flavor neutrino oscillations.

To allow for comparisons with searches for $\nu_s$'s in other channels, we adopt a minimal ``3+1'' extension~\cite{ref:3plus1,ref:3plus1_2,ref:3plus1_3,ref:3plus1_4,ref:3plus1_5} of the three-flavor neutrino model by augmenting the neutrino state basis set with one sterile state.  The resulting mixing matrix can be parametrized as $U\nobreak=\nobreak R_{34}S_{24}S_{14}R_{23}S_{13}R_{12}$ \cite{ref:matrixpram}, where $R_{ij}$ represents a rotation by the mixing angle $\theta_{ij}$, and $S_{ij}$ represents a complex rotation by the mixing angle $\theta_{ij}$ and the $CP$-violating phase $\delta_{ij}$. This model introduces additional parameters compared to the three-flavor model:  three new mixing angles ($\theta_{14}$, $\theta_{24}$, and $\theta_{34}$), two $CP$-violating phases ($\delta_{14}$ and $\delta_{24}$), and three new mass splittings, with only one being independent. In this analysis, we express the oscillation probabilities in terms of $\Delta m^2_{41}$.

The functional form for the NC disappearance probability can be illustrated by the approximate expression~\cite{ref:minos},
\begin{eqnarray}\label{eq:nc}
1 - P(\numu \to \nus) & \approx & 1 - \frac{1}{2}\cos^4\theta_{14} \cos^2\theta_{34} \sin^2 2\theta_{24} \nonumber \\*
& & + A \sin^2 \Delta_{31} - B \sin 2\Delta_{31},
\end{eqnarray}
where $\Delta_{31}$ = $\frac{\Delta m^2_{31}L}{4E}$. The 1/2 factor in the second term results from rapid oscillations driven by $\Delta m^2_{41}$, which average out at the FD due to our limited detector energy resolution~\cite{ref:pcSP}. The terms $A$ and $B$ are functions of the mixing angles and phases. To first order, $A=\sin^2\theta_{34}\sin^2 2\theta_{23}$ and $B=\frac{1}{2}\sin\delta_{24}\sin\theta_{24} \sin 2\theta_{34} \sin 2\theta_{23}$. The NC sample is therefore sensitive to $\theta_{24}$, $\theta_{34}$, and $\delta_{24}$. 
We perform a counting experiment comparing the FD NC rate to unoscillated and oscillated predicted rates that is valid for $0.05\leq \Delta m^2_{41}\leq 0.5$\,eV$^2$. In this range, the analysis is not sensitive to oscillations affecting the rates in the ND, present at larger $\Delta m^2_{41}$ values. Within the same range, the analysis is also insensitive to degenerate solutions with the three-flavor model, occurring when $\Delta m^2_{41}\simeq\Delta m^2_{32}$. 
Using an exact formulation of the 3+1 model that includes matter effects, we fit the data for $\theta_{24}$ and $\theta_{34}$ using the same oscillation parameter values and uncertainties as for the three-neutrino oscillation prediction, and profile over values of $\delta_{24}$. 
We estimate parameters by minimizing the expression,
\begin{equation}\label{eq:stat}
\resizebox{.9\hsize}{!}{$\chi^2 = 2\left( {F^{\rm{pred}}}- F^{\rm{data}} +F^{\rm{data}}\ln\frac{F^{\rm{data}}}{F^{\rm{pred}}} \right) +  \sum_{i} \left(\frac{\Delta U_i}{\sigma_{U_i}}\right)^2$}
\end{equation}

\begin{figure}[!t]
\includegraphics[scale=0.45]{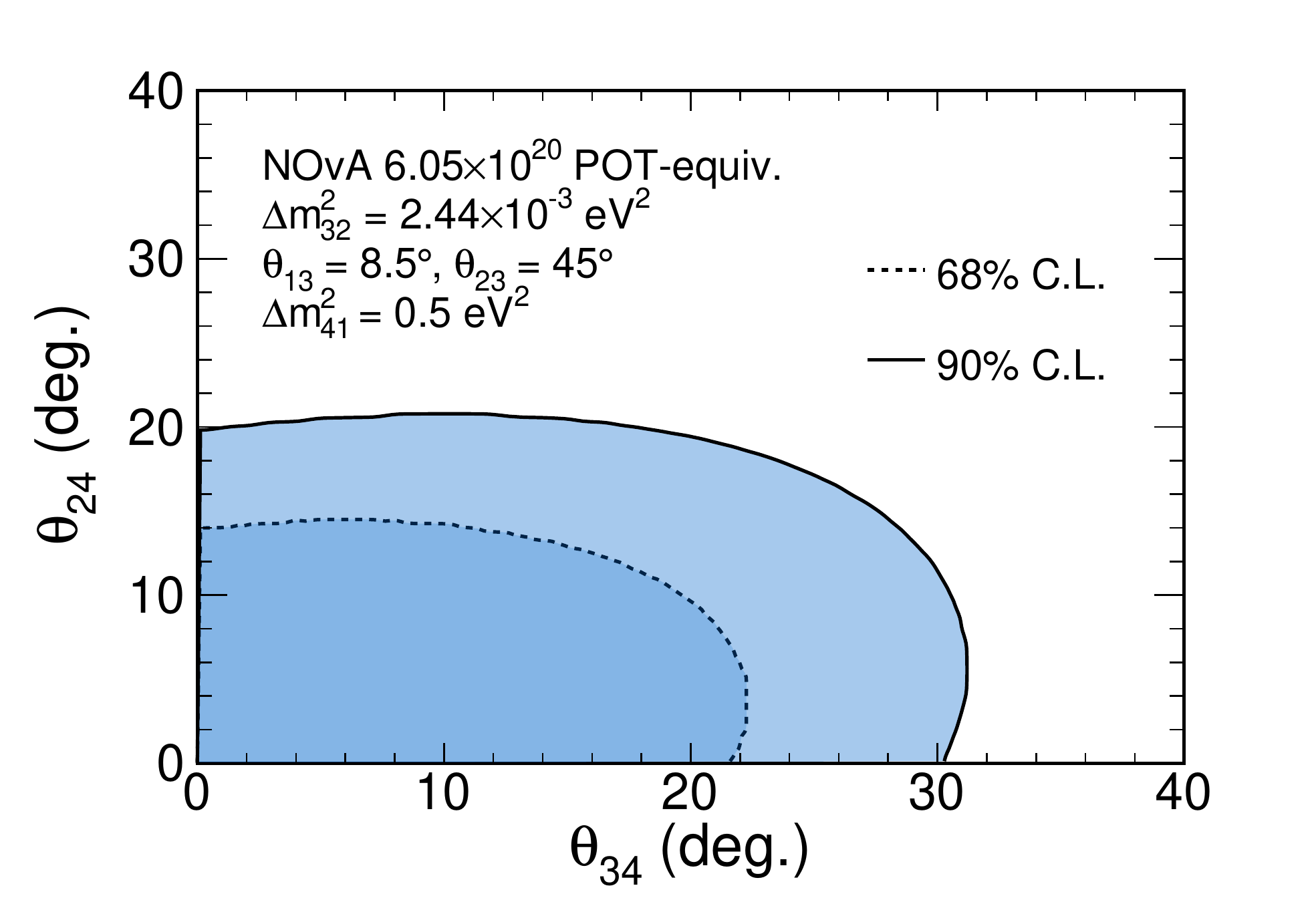}
\includegraphics[scale=0.45]{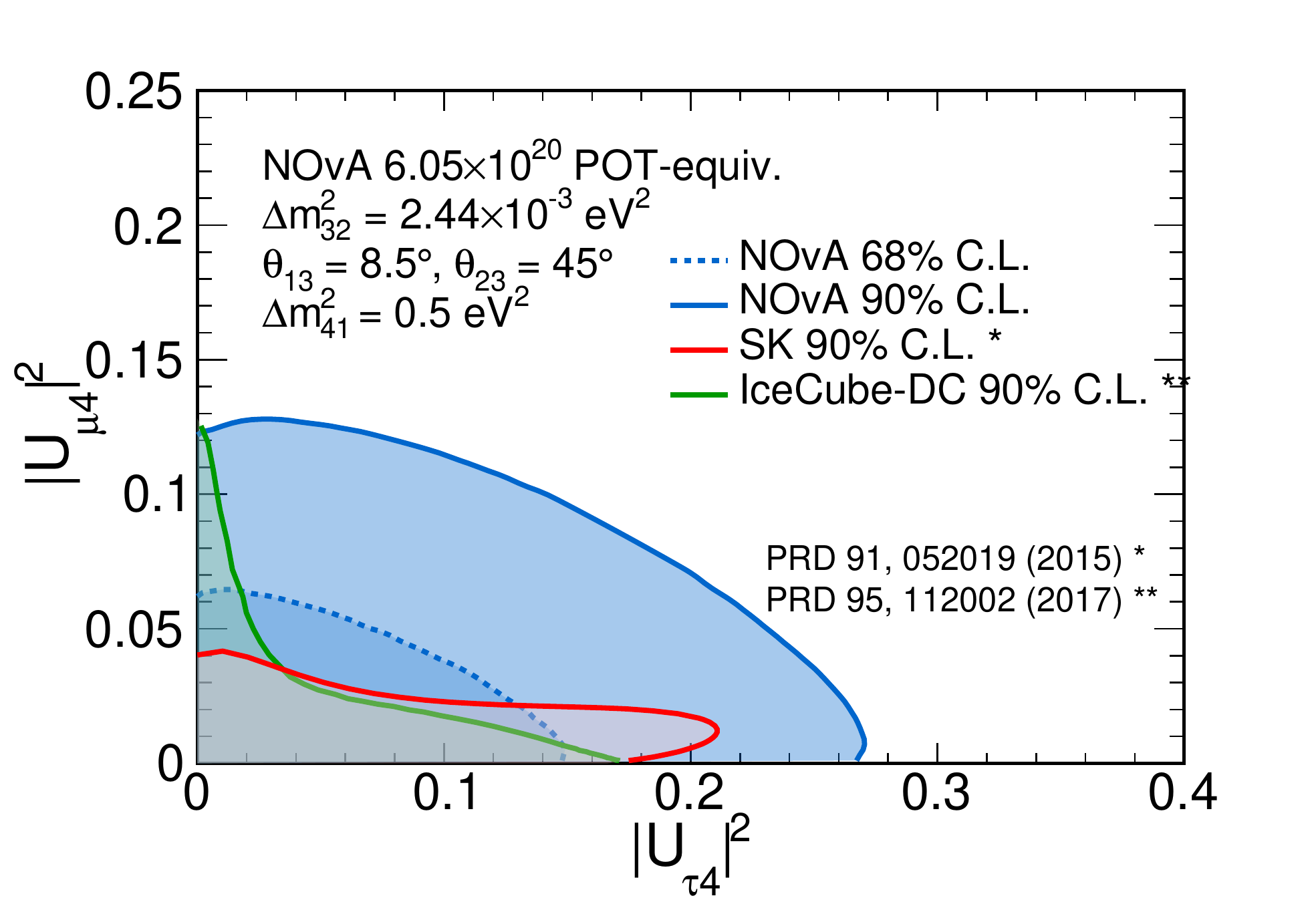}
\captionsetup{justification=RaggedRight}
\caption{Top: The 68\% (dashed) and 90\% (solid) Feldman-Cousins nonexcluded regions (shaded) for the mixing angles $\theta_{24}$ and $\theta_{34}$. Bottom: The 68\% (dashed) and 90\% (solid) Feldman-Cousins nonexcluded regions (shaded) in terms of $|U_{\mu4}|^{2}$ and $|U_{\tau4}|^{2} $ where we assume $\cos^2\theta_{14}=1$ in both cases.} 
\label{fig:results}
\end{figure}

The expected number of events is varied as a function of the oscillation parameters and of Gaussian-distributed penalty terms controlling the systematic uncertainties $U_{i}$. 
For the $i$th systematic uncertainty, $\Delta U_i$ denotes the amount the best fit is shifted by, and $\sigma_{U_i}$ denotes one standard deviation. The effects of each systematic uncertainty on the mixing angle measurement are summarized in Table~\ref{tab:SystSummary}.
Using the Feldman-Cousins unified approach~\cite{ref:feldman-cousins}, we compute $68\%$ and $90\%$ confidence levels resulting in the nonexcluded regions shown in Fig.~\ref{fig:results}.

For the 3+1 model, limits of $\theta_{24}$\,$<$\,$20.8^{\circ}$ and $\theta_{34}$\,$<$\,$31.2^{\circ}$ are obtained at the 90\% C.L. If expressed in terms of the relevant matrix elements
\begin{align}
|U_{\mu4}|^2 =&\,\,\cos^2\theta_{14}\sin^2\theta_{24} \\
|U_{\tau4}|^2= & \,\,\cos^2\theta_{14}\cos^2\theta_{24}\sin^2\theta_{34},
\label{eq:DisapToApp}
\end{align}
these limits become $|U_{\mu4}|^{2}$\,$<$\,0.126 and $|U_{\tau4}|^{2}$\,$<$\,0.268 at the 90\% C.L., where we conservatively assume $\cos^2\theta_{14}$\,=\,1 in both cases. This analysis is not sensitive to $\theta_{14}$ which is constrained to be small by reactor experiments~\cite{ref:Mention}. A comparison with present world-leading limits on $\theta_{34}$, $\theta_{24}$, $|U_{\mu4}|^{2}$, and $|U_{\tau4}|^{2}$ is shown in Table~\ref{tab:limits}.

\begin{table}[!ht]
\captionsetup{justification=RaggedRight}
  \caption{The 90\% C.L. upper limits on sterile mixing angles and matrix elements for NOvA compared to MINOS~\cite{ref:minos}, Super-Kamiokande~\cite{ref:superk}, IceCube~\cite{ref:IceCube}, and IceCube-DeepCore~\cite{ref:DeepCore}. The limits are shown for $\Delta m^2_{41} = 0.5$~eV$^2$ for all experiments, except for IceCube-DeepCore, where the results are reported for $\Delta m^2_{41} = 1.0$~eV$^2$.}
  \centering
  \begin{tabular}{c c c c c }
    \hline\hline
    & $\theta_{24}$ & $\theta_{34}$ & $|U_{\mu4}|^2$ & $|U_{\tau4}|^2$   \\
    \hline
    NOvA  & $20.8^{\circ}$ & $31.2^{\circ}$ & 0.126 & 0.268  \\
    MINOS & $7.3^{\circ}$ & $26.6^{\circ}$ & 0.016 & 0.20  \\
    SuperK & $11.7^{\circ}$ & $25.1^{\circ}$ & 0.041 & 0.18  \\
    IceCube & $4.1^{\circ}$ & \-- & 0.005 & \--   \\
    IceCube-DeepCore & $19.4^{\circ}$ & $22.8^{\circ}$ & 0.11 & 0.15  \\
    \hline\hline
  \end{tabular}%
    
  \label{tab:limits}
\end{table}

In conclusion, with an exposure of 6.05$\times$10$^{20}$\,POT-equivalent, we observe 95 NC-like events in the FD, compared with an expectation of ${83.5\,\pm\,9.7\mbox{(stat.)} \pm 9.4} \mbox{(syst.)}$. This result is consistent with three-flavor mixing within $1.03~\sigma$. No evidence for depletion of NC
events is observed in the FD at a distance of 810\,km from the neutrino source and NOvA sees no evidence for $\nu_s$ mixing. We set limits of $\theta_{24}$\,$<$\,$20.8^{\circ}$ and $\theta_{34}$\,$<$\,$31.2^{\circ}$ in a 3+1 model scenario.

Looking forward,  an overall fourfold increase in beam exposure is expected over the life of the experiment, which by itself will enable NOvA to be competitive with current experimental bounds on $\theta_{34}$. In addition, NOvA is implementing improvements in NC identification and in cosmogenic background rejection, working to reduce systematic uncertainties, and to include effects due to $\nu_s$ oscillations in the ND, further increasing the sensitivity of sterile neutrino probes over an extended $\Delta m_{14}^2$ range.

The NOvA collaboration uses the resources of the Fermi National Accelerator Laboratory (Fermilab), a U.S. Department of Energy, Office of Science, HEP User Facility. Fermilab is managed by Fermi Research Alliance, LLC (FRA), acting under Contract No. DE-AC02-07CH11359. This research was supported by the U.S. Department of Energy; the U.S. National Science Foundation; the Department of Science and Technology, India; the European Research Council; the MSMT CR, GA UK, Czech Republic; the RAS, RMES, and RFBR, Russia; CNPq and FAPEG, Brazil; and the State and University of Minnesota. We are grateful for the contributions of the staff at Fermilab and the NOvA Far Detector Laboratory.

\bibliographystyle{apsrev4-1}
\bibliography{nova_nc}

\end{document}